\documentclass[aps,prd,twocolumn,amsmath,superscriptaddress,amssymb,showpacs,floatfix,nofootinbib,longbibliography]{revtex4-1}

\usepackage{graphicx}
\usepackage{dcolumn}
\usepackage{xcolor}
\usepackage{bm}
\usepackage{natbib}
\usepackage{calligra}
\usepackage[T1]{fontenc}
\usepackage{egothic}
\usepackage[T1]{fontenc}
\newfont{\rsfsten}{rsfs10 scaled 1200}
\newfont{\rsfsseven}{rsfs10 scaled 1200}
\newfont{\rsfsfive}{rsfs10 scaled 1200}
\usepackage{epsfig}
\usepackage{units}
\usepackage[utf8]{inputenc}
\usepackage{hyperref}

\newcommand{\aap}{{Astron.~Astrophys.}}
\newcommand{\apjl}{{Astrophys.~J.~Lett.}}
\newcommand{\apjs}{{Astrophys.~J.~Supp.}}
\newcommand{\aj}{{Astron.~J.}}
\newcommand{\mnras}{{Mon.~Not.~R.~Astron.~Soc.}}

\newcommand{\be}{\begin{equation}}
\newcommand{\ee}{\end{equation}}
\newcommand{\bea}{\begin{eqnarray}}
\newcommand{\eea}{\end{eqnarray}}

\def\lsim{\mathrel{\raise.3ex\hbox{$<$\kern-.75em\lower1ex\hbox{$\sim$}}}}
\def\gsim{\mathrel{\raise.3ex\hbox{$>$\kern-.75em\lower1ex\hbox{$\sim$}}}}

\begin{document}

\title{Possible counterpart signal of the Fermi bubbles at the cosmic-ray positrons}

\author{Ilias Cholis}
\email{cholis@oakland.edu, ORCID: orcid.org/0000-0002-3805-6478}
\affiliation{Department of Physics, Oakland University, Rochester, Michigan, 48309, USA}
\author{Iason Krommydas}
\email{ik23@rice.edu, ORCID: orcid.org/0000-0001-7849-8863}
\affiliation{Physics Division, National Technical University of Athens, Zografou, Athens, 15780, Greece}
\affiliation{Department of Physics and Astronomy, Rice University, Houston, Texas, 77005, USA}
\date{\today}

\begin{abstract}
The inner galaxy has hosted cosmic-ray burst events including those responsible for the gamma-ray
\textit{Fermi} bubbles and the \textit{eROSITA} bubbles in X-rays. In this work, we study the 
\textit{AMS-02} positron fraction and find three features around 12, 21 and 48 GeV of which the
lowest energy has a 1.4 to 4.9-$\sigma$ significance, depending on astrophysical background assumptions. 
Using background simulations that explain the cosmic-ray positron fraction, positron flux and electron plus positron flux, 
by primary, secondary cosmic rays and cosmic rays from local pulsars, we test these spectral features as originating from 
electron/positron burst events from the inner galaxy. We find the 12 GeV feature, to be explained by an event of age 
$\tau \simeq 3 - 10$ Myr; in agreement with the proposed age of the \textit{Fermi} bubbles. Furthermore, 
the energy in cosmic-ray electrons and positrons propagating along the galactic disk and not within the \textit{Fermi} bubbles
volume, is estimated to be $10^{51.5}-10^{57.5}$ ergs, or $O(10^{-4}) -O(1)$ the cosmic-ray energy causing the 
\textit{Fermi} bubbles. We advocate that these positron fraction features, are the counterpart signals of the 
\textit{Fermi} bubbles, or of substructures in them, or of the \textit{eROSITA} bubbles.
\end{abstract}

\maketitle

The \textit{Fermi} bubbles \cite{Su:2010qj, Fermi-LAT:2014sfa}, discovered in 
\textit{Fermi} Large Area Telescope (\textit{Fermi}-LAT) \cite{Gehrels:1999ri, fermiURL} gamma-ray 
observations, have finite size and well defined limits. This suggests their origin 
is cosmic-ray activity from the inner galaxy, possibly associated to its
supermassive black hole. This activity may leptonic \cite{Crocker:2010qn, Cheng:2011xd, 
Mertsch:2011es, Guo:2011eg, Guo:2011ip, Yang:2012fy, Carretti:2013sc, Lacki:2013zsa} or hadronic
\cite{Crocker:2010qn, Crocker:2010dg, Lacki:2013zsa, Crocker:2014fla}. Moreover, these bubbles 
may be connected to the galactic center excess in gamma rays \cite{Goodenough:2009gk, 
Hooper:2010mq, Abazajian:2010zy, Hooper:2011ti, Hooper:2013rwa, Gordon:2013vta, Daylan:2014rsa, 
Calore:2014xka, Zhou:2014lva, Fermi-LAT:2015sau, Huang:2015rlu, Linden:2016rcf, DiMauro:2021raz, 
Cholis:2021rpp}, by both excesses resulting from successive cosmic-ray bursts 
\cite{Petrovic:2014uda, Carlson:2014cwa, Cholis:2015dea}. Additionally, the observation of 
``cocoons'' in the southern \cite{Su:2010qj, Fermi-LAT:2014sfa} and the northern \cite{Balaji:2018rwz}
galactic hemispheres within the \textit{Fermi} bubbles, and of bubbles in keV 
X-rays by the \textit{eROSITA} telescope \cite{Predehl:2020kyq}, suggests that the inner galaxy undergoes 
episodic events of enhanced cosmic-ray injection. Furthermore, \textit{ROSAT} and \textit{Suzaku} X-ray 
observations found related signals of the \textit{Fermi} bubbles \cite{1997ApJ...485..125S,2013ApJ...779...57K, 
Tahara:2015mia}, with connections to microwave observations as well \cite{2003ApJ...582..246B}. Finally, 
a similar excess by morphology was known as the \textit{WMAP} haze \cite{Finkbeiner:2003im, 
Dobler:2007wv}. This is likely the microwave counterpart signal of the \textit{Fermi} bubbles
 \cite{Dobler:2009xz}; and was confirmed by \textit{Planck} \cite{Dobler:2012ef, Planck:2012opn, 
 Planck:2015ica}. All these discoveries, point to the \textit{Fermi} bubbles coming from the inverse 
 Compton scattering (ICS) of cosmic-ray electrons ($e^{-}$) and positrons ($e^{+}$) \cite{Crocker:2010qn, Cheng:2011xd, 
 Mertsch:2011es, Guo:2011eg, Guo:2011ip, Yang:2012fy, Carretti:2013sc, Yang:2013kca, Lacki:2013zsa, Yang:2017tjr}. 

If positrons are contributing to the bubbles emission,  a fraction of the originally injected positrons could escape 
the bubbles regions that expand perpendicularly to the galactic disk. These cosmic-ray positrons would reach us, 
contributing to the local positron flux. In this \textit{letter} we claim a hint of that event leading to 
a feature in the positron fraction around energies of 12 GeV (see also \cite{Cholis:2021kqk}). We also find a 
second feature at 21 GeV, that could come from a more recent burst event from the inner galaxy. We propose 
that these two features in positrons are associated in energy and age to either the \textit{eROSITA} 
and \textit{Fermi} bubbles, or the  \textit{Fermi} bubbles and the cocoons within them.

A clear rise of the positron fraction ($e^{+}/(e^{+}+e^{-})$) above 5 GeV has been discovered the Payload for 
Antimatter Matter Exploration and Light-nuclei Astrophysics (\textit{PAMELA}) and the Alpha Magnetic 
Spectrometer (\textit{AMS-02}) \cite{PAMELA:2013vxg, AMS:2014bun, AMS:2019iwo, AMS:2021nhj}.  
That rise is in tension with expectations from positrons produced in inelastic collisions of 
cosmic-ray protons and nuclei with the interstellar medium (ISM) gas \cite{Moskalenko:2001ya, Kachelriess:2015wpa, 
GALPROPSite, Strong:2015zva, Evoli:2008dv, DRAGONweb, Evoli:2011id, Pato:2010ih}, and indicates 
local sources of high-energy positrons.  Those could be near-by pulsars \cite{1987ICRC....2...92H, 1995PhRvD..52.3265A, 
1995A&A...294L..41A, Hooper:2008kg, Yuksel:2008rf, Profumo:2008ms, Malyshev:2009tw, Kawanaka:2009dk, 
Grasso:2009ma, 2010MNRAS.406L..25H, Linden:2013mqa, Cholis:2013psa, Yuan:2013eja, Yin:2013vaa, 
Cholis:2018izy, Evoli:2020szd, Manconi:2021xom, Orusa:2021tts, Cholis:2021kqk}, local and recent supernova 
remnants (SNRs) \cite{Blasi:2009hv, Mertsch:2009ph, Ahlers:2009ae, Blasi:2009bd, Kawanaka:2010uj, 
Fujita:2009wk, Cholis:2013lwa, Mertsch:2014poa, DiMauro:2014iia, Kohri:2015mga, Mertsch:2018bqd} 
(see however  \cite{Cholis:2013lwa, Mertsch:2014poa, Cholis:2017qlb, Tomassetti:2017izg}) or particle dark 
matter \cite{Bergstrom:2008gr, Cirelli:2008jk, Cholis:2008hb, Cirelli:2008pk, Nelson:2008hj, ArkaniHamed:2008qn, 
Cholis:2008qq, Cholis:2008wq, Harnik:2008uu, Fox:2008kb, Pospelov:2008jd, MarchRussell:2008tu, 
Chang:2011xn, Cholis:2013psa, Dienes:2013xff, Finkbeiner:2007kk, Kopp:2013eka, Dev:2013hka, 
Klasen:2015uma, Yuan:2018rys, Sun:2020dla}.

\begin{figure}
\begin{centering}
\hspace{-0.6cm}
\includegraphics[width=3.72in,angle=0]{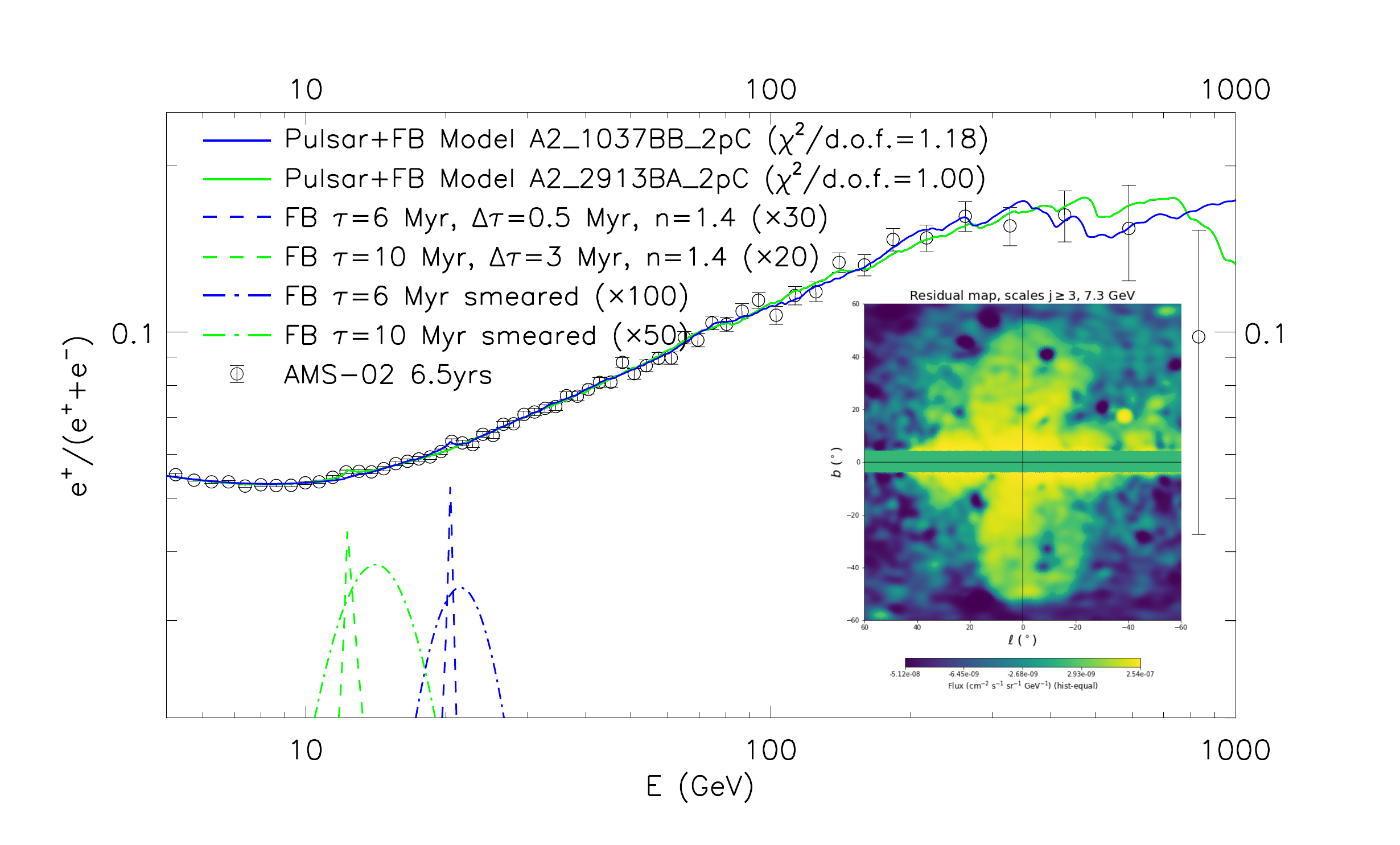}
\end{centering}
\vspace{-0.9cm}
\caption{The \textit{AMS-02} positron fraction. 
Each model contains the full astrophysical background contributions and that coming from inner galaxy burst events 
(dashed lines). The total positron fraction is in solid lines. The dot-dashed lines show the effect of propagation-induced
smearing of the burst event's contribution. These examples assume same ISM 
propagation conditions but different energy outputs, ages $\tau$ and durations $\Delta\tau$ for the $e^{\pm}$ 
burst events. Features in the modeled lines at energies $\gsim 100$ GeV originate from 
powerful local pulsars. The embedded figure shows the gamma-ray \textit{Fermi} bubbles from \cite{Balaji:2018rwz}.}
\vspace{-0.6cm}
\label{fig:PF_Features}
\end{figure}

In this work, we assume that local pulsars are responsible for the observed cosmic-ray positron flux and fraction 
measurements following \cite{Cholis:2021kqk}. Pulsars are energetic sources with their properties probed from radio 
wavelengths to TeV gamma-rays \cite{1968Natur.217..709H, 1998MNRAS.301..235G, 1999ApJS..123..627S, 
Weisberg_1999, Everett:2000yj, McLaughlin:2003zz, Weisberg:2003ud, 2005AJ....129.1993M, 2009A&A...504..525S, 
2012A&A...544A.100M, 2012A&A...540A..28D, 2013MNRAS.433.3325S, 2019A&A...629A.140S, 2009A&A...504..525S, 
2012A&A...544A.100M, 2012A&A...540A..28D, 2013MNRAS.433.3325S, 2019A&A...629A.140S,  2019ApJ...871..246M, 
2021NatAs...5..552A, 1971ApJ...167L..67G, Becker:2002wf, Gentile:2013yka, 2013MNRAS.433.3325S, 
Guillot:2019vqp, Zhao:2021ilq, 1994ApJS...90..789U, 2010ApJS..187..460A, Fermi-LAT:2013svs, Buhler:2013zrp, 
Cholis:2014noa, Fermi-LAT:2019yla, 2009ApJ...700L.127A, MAGIC:2015ggt, HESS:2017lee, Abeysekara:2017hyn, 
HAWC:2019tcx}. Moreover, they may produce multiple spectral features at high energies \cite{Malyshev:2009tw, 
Cholis:2017ccs}  that dark matter or SNRs may not. Hypothesizing local pulsars being responsible for the 
overall rise of the positron fraction spectrum, is the conservative assumption when claiming that spectral features 
in the data are associated to activity from the inner galaxy, instead of coming from an old and energetic 
local pulsar. In Fig.~\ref{fig:PF_Features}, we show the positron fraction, fitted by two models 
that each explains the features at 12 and 21 GeV respectively. 

We use the \textit{AMS-02} cosmic-ray positron fraction \cite{AMS:2019iwo}, -where systematics 
as instrumental exposure ones may cancel out-, to test the significance of spectral features and 
constrain the properties of cosmic-ray bursts.
We have tested the positron flux \cite{AMS:2019rhg}, and find the features to be present there as well; 
with however, the cosmic-ray bursts properties be less well constrained. Thus, we focus on
positron fraction.

Defining the 12 and 21 GeV features as signal, the backgrounds from 
Ref.~\cite{Cholis:2021kqk}, were fit to the \textit{AMS-02} positron fraction \cite{AMS:2019iwo}, 
positron flux \cite{AMS:2019rhg} and total $e^{-} + e^{+}$ flux 
\cite{AMS:2021nhj}; and the \textit{DAMPE} and \textit{CALET} total $e^{-} + e^{+}$ fluxes \cite{DAMPE:2017fbg, 
Adriani:2018ktz}. Those background simulations include primary electrons accelerated by conventional 
cosmic-ray sources as SNRs, secondary electrons and positrons produced in inelastic collisions between 
cosmic-ray protons and nuclei with the ISM gas and local galactic pulsars~\cite{Cholis:2021kqk}. The 
cosmic-ray propagation, spatial distribution of sources and averaged 
injection spectra of cosmic-ray nuclei assumptions are in agreement with the \textit{AMS-02} observations 
\cite{Cholis:2021rpp}, and compatible with~\cite{Trotta:2010mx, Pato:2010ih}. 

\textit{Our simulations} of~\cite{Cholis:2021kqk}, include $(5-18)\times 10^{3}$ unique pulsars within 4 
kiloparsec (kpc) from the Sun. These backgrounds, model uncertainties on a) the stochastic nature of the neutron 
stars' birth in space and time, with each pulsar having a unique combination. We assumed a total birth 
rate of 0.6-2 pulsars per century for the Milky Way~\cite{1999MNRAS.302..693D, Vranesevic:2003tp, 
FaucherGiguere:2005ny, Keane:2008jj}, and relied on observations and modeling of the pulsars' spatial 
distribution from \cite{Manchester:2001fp, FaucherGiguere:2005ny, Lorimer:2003qc,Lorimer:2006qs}. 
Our simulations also account for b) each pulsar having a unique initial spin-down power, following a 
distribution that while uncertain is constrained by 
radio observations \cite{FaucherGiguere:2005ny, Manchester:2004bp, ATNFSite}. Furthermore, we
account for c) the pulsars' time-evolution uncertainties, by testing distinct values for all pulsars' braking index $\kappa$, 
and characteristic spin-down timescale $\tau_{0}$. 
The combination of $\kappa$ and  $\tau_{0}$ values simulated where in agreement with 
expected surface magnetic field and period distributions from~\cite{FaucherGiguere:2005ny}.  
Another modeled uncertainty is d) the fraction $\eta$ of the pulsars' total rotational energy to cosmic-ray 
$e^{\pm}$ injected to the ISM, and the relevant injection spectral index $n$. 
Each pulsar has a unique set of $\eta$ and $n$, following alternative choices for a log-normal distribution
and a uniform distribution respectively. The spectra from all pulsars have an exponential cut-off at 
10 TeV (see Ref.~\cite{Cholis:2021kqk} for details). 
Finally, these background simulations accounted for e) cosmic-ray propagation uncertainties through the ISM 
and the Heliosphere. For the ISM propagation, the most significant assumptions are those associated to their 
diffusion and their energy losses. For isotropic and homogeneous conditions, diffusion is described by a rigidity 
($R$)-dependent coefficient, $D(R) = D_{0} (R / (1 \, GV))^{\delta}$, where $D_{0}$ is the normalization at 1 GV 
and $\delta$ is the diffusion index \cite{1941DoSSR..30..301K, 1967PhFl...10.1417K}. We propagate cosmic rays 
within a cylinder of radius 20 kpc of half-height $z_{L}$, centered at the galactic center. 
For cosmic-ray $e^{\pm}$ of $O(10)-O(100)$ GeV, ICS and synchrotron radiation
dominate the energy losses. At these energies for ICS the Thomson cross-section \cite{1929ZPhy...52..853K} 
approximates the Klein-Nishina well \cite{1970RvMP...42..237B}. Thus, the energy-loss rate 
scales as, $dE/dt = -b (E/(1 \, GeV ))^{2}$. The parameter $b$, scales with the energy density in the CMB 
and interstellar radiation field photons and the local galactic magnetic field. Diffusive reacceleration 
\cite{1994ApJ...431..705S} and convective winds affect mostly the cosmic-ray nuclei relevant for the calculation 
of cosmic-ray secondaries and are included. We use models ``A'', ``C'' and ``E'' from Refs.~\cite{Cholis:2021kqk} 
and \cite{Cholis:2021rpp} that fit the observed \textit{AMS-02} cosmic-ray hydrogen, helium, carbon, oxygen fluxes 
and the beryllium-to-carbon, boron-to-carbon and oxygen-to-carbon ratios. For the secondary fluxes \path{GALPROP} v54 
\cite{galprop, GALPROPSite}, has been used. 
We use three choices for the averaged ISM energy losses burst $e^{\pm}$ experience in traveling from the galactic 
center the Sun.
Those are denoted as ``2'', ``4'' and ``5''  \footnote{We retain and extend the notation of~\cite{Cholis:2021kqk} for easier reference.}. These ISM models are given in Table~\ref{tab:ISMBack}. 
For the local pulsars contributing to the background, only option ``2'' with $b=8.02 \times 10^{-6}$ 
GeV$^{-1}$ kyrs$^{-1}$ is used, representing the upper limit of local energy-loss uncertainties \cite{galprop, GALPROPSite, Porter:2017vaa}. As we move closer to the galactic center, both the 
radiation field and the magnetic field amplitude increase (on average). Thus, assumptions of higher energy losses than 
the local ones are required to study the burst events. Options ``4'' and ``5'' model higher energy losses (than ``2'') 
and are explicitly used to model cosmic rays originating from the inner galaxy bursts. All cosmic-ray spectra are affected by solar modulation. We implement 
the time-, charge- and rigidity-dependent formula for the solar modulation potential from \cite{Cholis:2015gna} 
following the allowed ranges on the $\phi_0$ and $\phi_1$ modulation parameters that describe it from
\cite{Cholis:2020tpi, Cholis:2022rwf}.

\begin{table}[t]
    \begin{tabular}{ccccc}
         \hline
           Model &  $z_{L}$ (kpc) & $b$ ($\times 10^{-6}$GeV$^{-1}$kyrs$^{-1}$) & $D_{0}$ (pc$^2$/kyr) & $\delta$\\
            \hline \hline
            A2 &  5.7 & 8.02 & 140.2 & 0.33 \\
            A4 &  5.7 & 16.04 & 140.2 & 0.33 \\
            A5 &  5.7 & 24.12 & 140.2 & 0.33 \\     
            C2 &  5.5 & 8.02 & 92.1 & 0.40 \\
            C4 &  5.5 & 16.04 & 92.1 & 0.40 \\
            C5 &  5.5 & 24.12 & 92.1 & 0.40 \\      
            E2 &  6.0 & 8.02 & 51.3 & 0.50 \\
            E4 &  6.0 & 16.04 & 51.3 & 0.50 \\
            E5 &  6.0 & 24.12 & 51.3 & 0.50 \\
        \hline \hline 
        \end{tabular}
        \vspace{-0.3cm}
\caption{The cosmic-ray propagation parameters for the ISM models that we use.} 
\vspace{-0.5cm}
\label{tab:ISMBack}
\end{table}

\textit{In fitting} the background models,
there are seven fitted parameters. There are three normalization factors, for primary $e^{-}$, secondary $e^{\pm}$, 
and pulsar $e^{\pm}$ fluxes. Two more parameters allow for the spectral
hardening or softening of the primary and the secondary cosmic rays, to model uncertainties in the ISM gas and the 
production efficiency and injection spectra of primary cosmic rays. Finally, 
the solar modulation parameters $\phi_0$ and $\phi_1$ are marginalized over. 
Once including the cosmic-ray burst component, an additional normalization is used to
set the overall energy deposited in cosmic-ray $e^{\pm}$, that escaped the inner galaxy along the 
galactic disk; assuming energy equipartition between the two species. 
We use a combination of \path{SciPy}'s \cite{2020SciPy-NMeth} \path{least_squares} routine from 
the \path{optimize} module and \path{iminuit} \cite{iminuit,James:1975dr} to fit our models (see Ref.~\cite{Cholis:2021kqk} for further details).

\textit{We test four basic properties for the cosmic-ray bursts}. These are the 
age $\tau$ of the event, defined from the start of the ejection event, and the duration of that event $\Delta \tau$, 
during which time $90\%$ of the energy was released. The third parameter is the total energy released into cosmic-ray 
$e^{\pm}$ \textit{escaping the Fermi bubbles, or cocoons, or eROSITA
bubbles volume} and reaching us. The forth parameter is the spectral index $n_{\textrm{burst}}$ of the released cosmic rays into the ISM, where,
\begin{equation}
\frac{dN}{dE}_{\textrm{burst}} \propto  E^{-n_{\textrm{burst}}} \textrm{Exp} \left\{ -\frac{E}{E_{\textrm{cut}}}\right\},
\label{eq:InjSpect}
\end{equation}
with $E_{\textrm{cut}} = 10$ TeV. As cosmic rays cool, the $\textit{AMS-02}$ data cannot probe the exact value 
of that exponential cut-off. We test values for $n_{\textrm{burst}}$ within $[1.4, 1.9]$. Thus, the total energy is 
set by the high end of the injected energy. The energy output of the burst can be probed by the 
cosmic-ray data, through the amplitude of that component as shown in Fig.~\ref{fig:PF_Features}.

While a burst has an intrinsic duration $\Delta \tau$, that gives a width 
to its positron fraction spectral feature, other effects can increase that width. The 
ICS is a stochastic process (see \cite{John:2022asa}). Also, cosmic rays
reaching us diffuse through unique paths. Thus, their travel time varies and they experience
unique energy losses as the energy-loss rate is position-dependent within the galaxy. 
To account for these additional propagation-induced smearing of the burst's flux, 
we follow Ref.~\cite{Malyshev:2009tw}, where the propagation-induced width $\Delta E$ 
of an otherwise $e^{\pm}$ with propagated energy
$E$ is $\Delta E/E \simeq 0.05 (E/ (1 \textrm{TeV}))^{-1/3}$. For a positron flux feature, this causes a slight
shift of its peak to higher energies, which for a positron fraction feature over a falling $e^{\pm}$ primary and 
secondary only background fluxes as those shown in Fig.~\ref{fig:PF_Features}, (by dot-dashed lines) is further enhanced. 
We consider the un-smoothed features to be a simplistic approximation, while the 
smeared ones a more realistic description of the burst's signal on the positron fraction. However, as we 
show the effect on the final information we derive from the positron fraction on the properties of the burst(s) is 
minimal. The exact prescription on the propagation smearing is of minor importance. 

\begin{figure}
\begin{centering}
\hspace{-0.6cm}
\includegraphics[width=3.72in,angle=0]{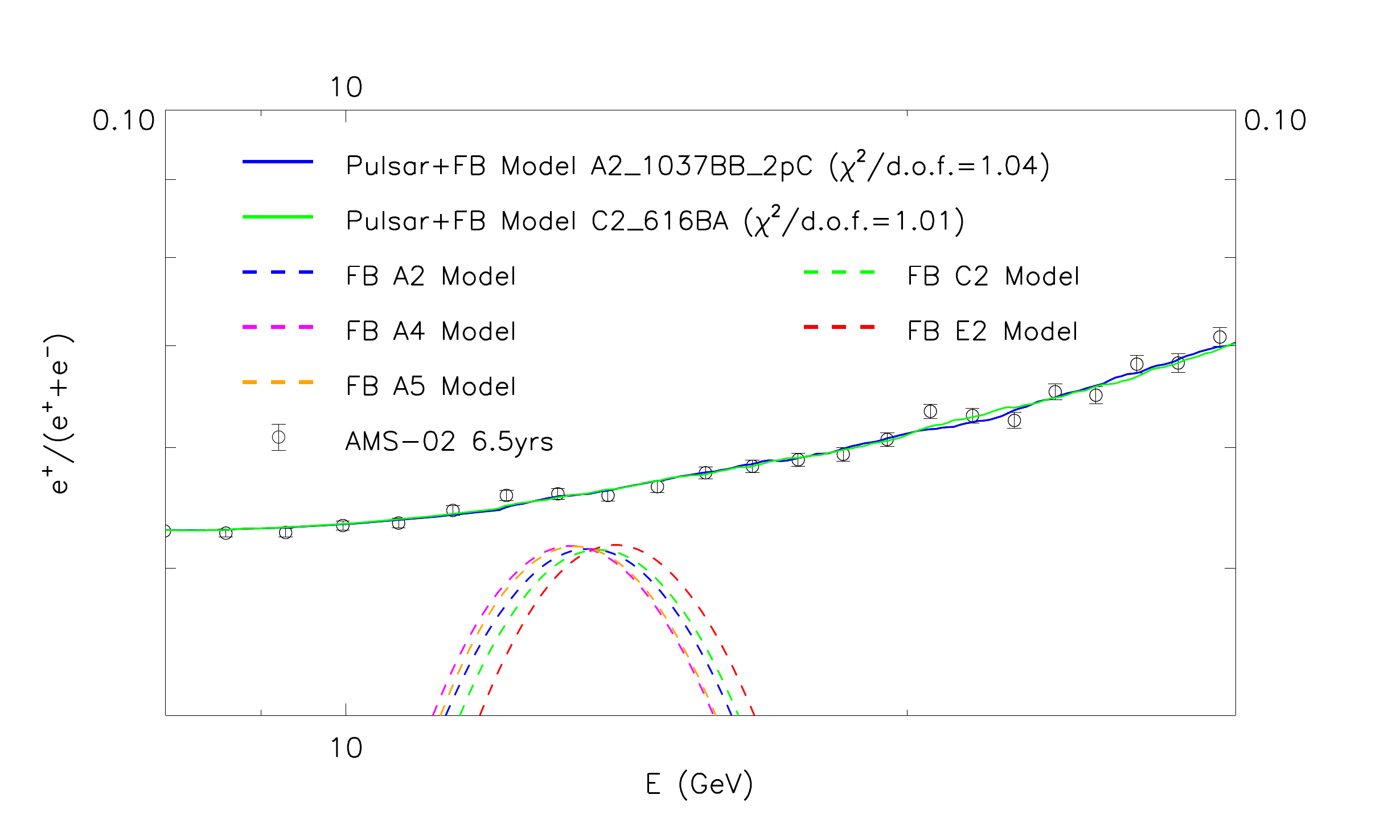}
\end{centering}
\vspace{-0.9cm}
\caption{
How different ISM assumptions affect the burst contribution to the positron fraction.
We take a burst duration $\Delta \tau = 10$ kyr and an injection index $n=1.4$. The burst's age for 
the ``A2'', ``C2'' and ``E2'' models is $\tau = 9.5$ Myr, while for the ``A4'' $\tau = 5$ Myr, and for "A5" 
$\tau = 3.4$ Myr.}
\vspace{-0.6cm}
\label{fig:PF_ISM_assump}
\end{figure}

In Fig.~\ref{fig:PF_ISM_assump}, we show the impact that different ISM assumptions have on the exact location 
of the peak and spectral shape of the burst. The lines have arbitrary normalizations. The normalization of the burst
signal changes as faster energy losses or diffusion will suppress the observed amplitude at the Sun. 
For ISM models ``2'', a burst event contributing at 12 GeV has $\tau \simeq 10$ Myr. 
Higher energy-loss assumptions from the inner galaxy to the Sun, move the 
feature's peak from a burst of a given age $\tau$ to lower energies. Thus, higher energy losses to those of the ISM models ``2'',
as expected for the inner galaxy will require bursts of younger age. For the feature at 12 GeV, that makes
the burst events to be $\tau \simeq 5$ Myr for the ``4'' and $\tau \simeq 3.5$ Myr  for the ``5'' ISM models.

\begin{figure*}
\begin{centering}
\hspace{-0.1cm}
\includegraphics[width=3.50in,angle=0]{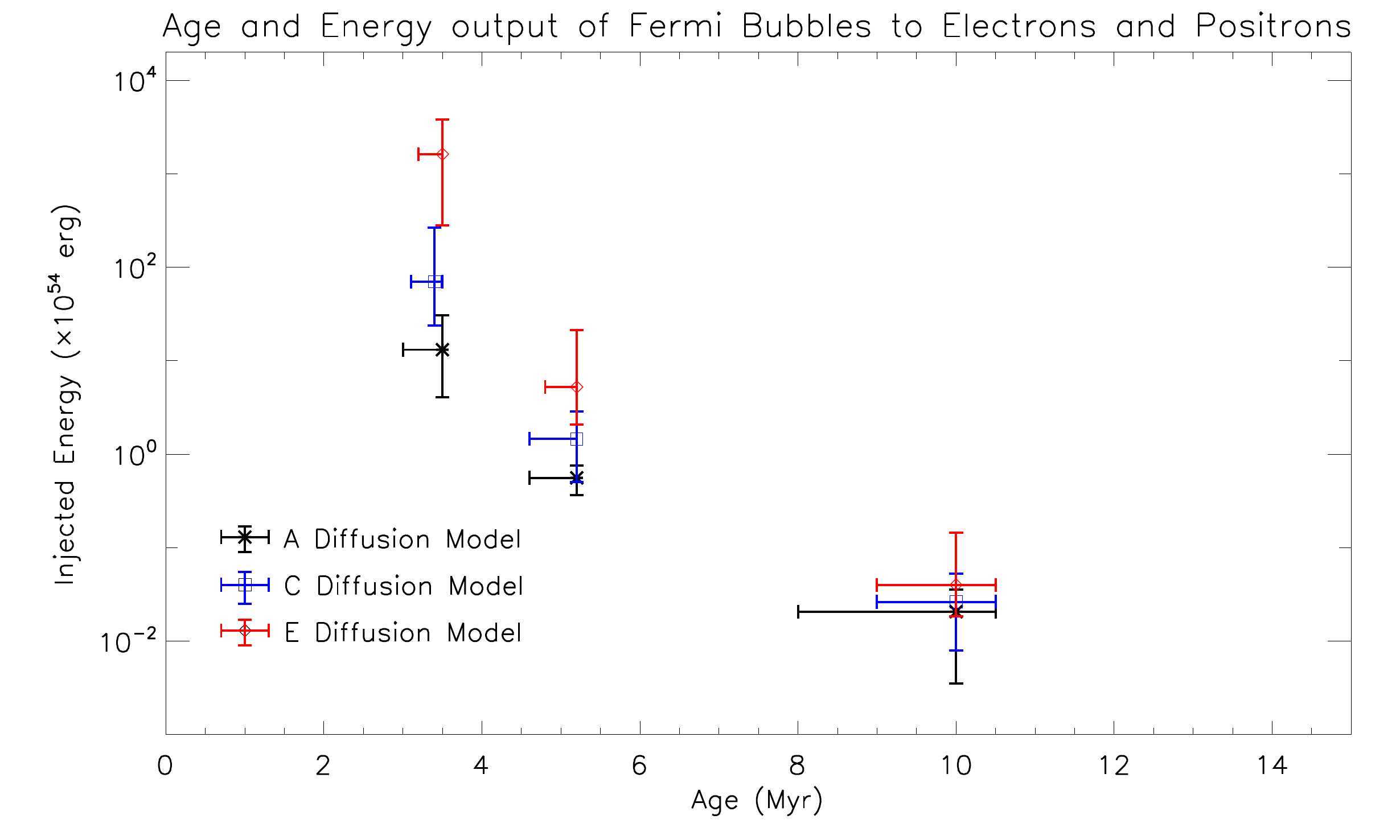}
\includegraphics[width=3.50in,angle=0]{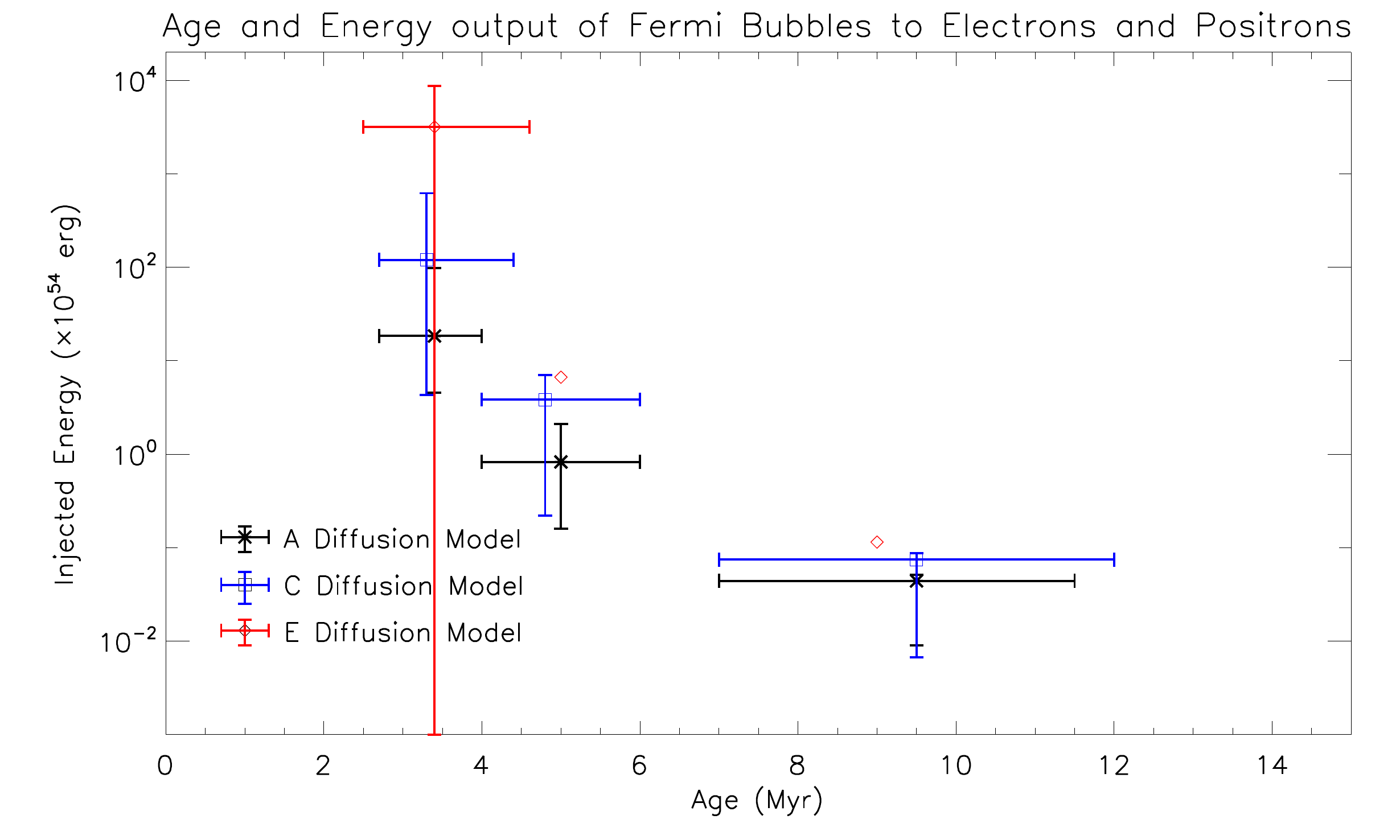}
\end{centering}
\vspace{-0.5cm}
\caption{Associated to the 12 GeV feature, the galactic center burst's parameter space inferred from fitting the positron fraction. 
We show the total injected energy in $e^{\pm}$ that escaped along the galactic disk and away from the bubbles regions 
versus the age of the burst. On the left panel we show results without the propagation-induced smearing, while on the right 
we test models that include this effect. The colors of the best fit points and $1 \sigma$ ranges (when statistically defined) are for 
different diffusion assumptions (``A'', ``C'' and ``E''), while the smallest ages are from the highest energy losses 
of ``5'' and the largest burst ages from the lowest energy losses of ``2'' (see Table~\ref{tab:ISMBack}).}
\vspace{-0.0cm}
\label{fig:12GeVFeature}
\end{figure*}

\textit{After fitting the galactic burst models}, we find a hint for at least one such event present in the
positron fraction spectrum. The 12 GeV feature is the most prominent with a statistical significance of 
$\Delta \chi^{2} \simeq 10$, with a $\Delta \chi^{2}$ range between 2 to 24 (i.e. 1.4 to 4.9 $\sigma$ significance) 
depending on the background model used. The 21 GeV feature has a significance of $\Delta \chi^{2}$ up to 6.1 
(2.4 $\sigma$); while a feature at 48 
GeV has a significance of $\Delta \chi^{2}$ up to 8.1 or 2.8 $\sigma$ \footnote{ 
The quoted $\Delta \chi^2$ values are accurate within $\pm 2$ depending on the minimization starting point.}.
Our burst simulations are tested with 20 
different background models from~\cite{Cholis:2021kqk}. Each background model gives a good fit to the positron 
fraction, the positron flux and the $e^{+}+e^{-}$ flux measurements. The bursts do not help explain energy ranges
beyond the features themselves.  Given these results, we focus on the properties of the burst associated with the 
most significant and robust 12 GeV feature. A burst event younger by a factor of 
0.6 and 1/4th would be responsible for the 21 and 48 GeV features respectively. In terms of its total energy we find the 
21 GeV feature to be roughly similar to the 12 GeV feature, which properties we describe next. 

In Fig.~\ref{fig:12GeVFeature}, we show for the 12 GeV feature the burst's age $\tau$ and $e^{\pm}$ injected energy parameter 
ranges. We show results where we ignore the propagation smearing (left panel) and where we include it
(right panel). This smearing does not significantly change the burst's inferred properties, with the ranges for
$\tau$ and the injected energy being marginally wider for the models with the smearing included. The other two 
burst tested parameters, i.e. $\Delta \tau$ and $n_{\textrm{burst}}$ are not constrained by the positron fraction. 
We tested values of  $\Delta \tau$  from 10 kyr (effectively instantaneous injection) and up to $\Delta \tau /\tau \simeq 1/3$. 
The presence of smearing reduces slightly the statistical significance of the feature. The $1 \sigma$ 
ranges are constructed based on the combination of all 20 background models, while the best-fit points show the best-fit model.

In this \textit{letter}, we used the cosmic-ray positron fraction observations from \textit{AMS-02} and find a statistically 
significant 12 GeV feature that can be explained by an inner galaxy burst event of age $\tau \simeq 3-10$ Myr and 
energy output to cosmic-ray $e^{\pm}$ escaping along the galactic disk (and reaching us) of $10^{51.5}$ to $10^{57.5}$ erg.
The larger ages of $\tau \simeq 10$ Myr come from fitting our low energy losses models ``2'', while more realistic 
assumptions for the inner galaxy (``4'' and ``5'') give $\tau \simeq 3-5$ Myr. This result is in exceptional
agreement with estimates for the age of \textit{Fermi} bubbles, as from simulations in \cite{Guo:2011eg, Guo:2011ip}. 
Furthermore, the \textit{Fermi} bubbles, if of leptonic origin, are expected to have originated by a burst of cosmic rays 
moving away from the disk with an energy of $10^{56} - 10^{57}$ erg. The 12 GeV feature could be the result of a
fraction of that cosmic-ray $e^{\pm}$ energy, escaping from the bubbles region and propagating along the galactic disk, 
thus reaching our location. The $10^{51.5}$ to $10^{57.5}$ erg is in agreement with that picture, with the ratio of cosmic-ray 
 energy along the disk to the energy within the bubbles being in the very approximate range of $O(10^{-4}) -O(1)$. 

There are two statistically less significant features at 21 and 48 GeV in the positron fraction.  
If astrophysical and from the inner galaxy, these features would be associated to more recent burst events. 
If the 12 GeV feature is the \textit{Fermi} bubbles' counterpart signal in the local positrons, then the smaller in 
size and later in time gamma-ray cocoons could be associated to one of those higher energy features. 
Alternatively, the \textit{eROSITA} bubbles with a similar energy output to the \textit{Fermi} bubbles but likely different 
underlying cosmic-ray spectrum could be the earlier event explaining the 12 GeV feature with the 21 GeV feature being 
connected to the \textit{Fermi} bubbles. A better understanding of the nature of these inner galaxy bursts detected in 
gamma rays and X rays will help constrain their properties. Also, with further cosmic-ray 
measurements, we will attain a better understanding on the local cosmic-ray $e^{\pm}$ fluxes and scrutinize these
 spectral features.

The background simulations from Ref.~\cite{Cholis:2021kqk}, account for primary $e^{-}$, secondary $e^{\pm}$ 
and $e^{\pm}$ from pulsars; and give good fits to the $e^{\pm}$ 
measurements from \textit{AMS-02}, \textit{DAMPE} and \textit{CALET}. They account for the stochastic nature 
of the neutron stars' birth in space and time, uncertainties on their birth rate, initial spin-down power, on the 
evolution of the pulsars' spin-down, on their injected $e^{\pm}$ fluxes and on cosmic-ray propagation uncertainties. 
Pulsars can also give spectral features at high energies. For an invisible in electromagnetic observations local pulsar within 
$\sim 0.5$ kpc, the 12 GeV feature would require it to be of age $\simeq 10$ Myr and with a high initial spin 
down power at least 3 times that of the Crab pulsar. We consider that an alternative explanation to be probed 
by multi-wavelength observations. 
In a separate paper \cite{Krommydas:2022}, we explore the possible impact these lower-energy features have on dark matter limits. 
We make publicly available our burst simulations in \cite{ZENODOFermiBubbles}, while our background simulations are available through \cite{Cholis:2021kqk, ZENODOBackgrounds}.
  
\textit{Acknowledgements:} 
We thank Bhaskaran Balaji, Patrick J. Fox, Dan Hooper and Samuel D. McDermott for useful discussions.
We acknowledge the use of \texttt{GALPROP} \cite{GALPROPSite, galprop}
and the \path{Python} \cite{10.5555/1593511} modules, \path{numpy} \cite{harris2020array},
\path{SciPy} \cite{2020SciPy-NMeth}, \path{pandas} \cite{reback2020pandas,mckinney-proc-scipy-2010}, \path{Jupyter} \cite{Kluyver2016jupyter}, and \path{iminuit} \cite{iminuit,James:1975dr}.
IC acknowledges support from the Michigan Space Grant Consortium, NASA Grant No. 80NSSC20M0124.
IC acknowledges that this material is based upon work supported by the U.S. Department of Energy, Office of Science, 
Office of High Energy Physics, under Award No. DE-SC0022352.
\bibliography{FermiBubbles_Positrons}

\end{document}